\documentclass[jkps,preprint,fleqn,showpacs,showkeys]{revtex4}
\usepackage[pdftex]{graphicx}
\usepackage{amssymb}
\usepackage{amsmath}
\usepackage{mathtools}
\usepackage{amsthm}
\usepackage{upgreek}
\usepackage{natbib}

\begin{document}
\setcounter{page}{0}
\title[]{The Bardeen-Petterson Effect as an Observable Support for the Blanford-Payne Process Black Hole Jet Production
}
\author{Uicheol Jang \surname{Jang}}

\email{temiy@cnu.ac.kr}
\thanks{}
\affiliation{Department of Astronomy, Space Science and Geology, Chungnam National Univ., Daejeon
 34134}
\author{Hongsu \surname{Kim}}
\affiliation{Center for Theoretical Astronomy , Korea Astronomy and Space Science Institute, Daejeon
34055}

\date[]{Received  - -}

\begin{abstract}
Relativistic jets are observed around accreting black holes, from stellar mass to supermassive black holes. However, their origin has not been fully understood. Although the Blanford-Payne process has been considered the most reliable theory for jet production, it has no observational proof or support. In this study, we suggest that the Bardeen-Petterson effect can provide observational evidence for the Blanford-Payne process. We studied black hole jet production using the Blanford-Payne process and the Bardeen-Peterson effect. As a result, we were able to calculate the observable timescale of black hole jet precession. 

\end{abstract}



\maketitle

\section{Introduction}
Accreting black holes are observed to produce relativistic jets. Although the production of active galactic nuclei (AGN) jets has been known as the Blanford-Payne process\cite{blandford1982hydromagnetic}, observational evidence or support has been rare. Therefore, we would like to suggest observable cases that can support observational evidence of the Blanford-Payne process. The Blanford-Payne process, as described, is a mechanism for producing AGN jets that are driven by gas (corona) pressure \cite{blandford1982hydromagnetic}. This interpretation shows that the jet outflow is closely related to the accretion disk. And the accretion disk around a spinning black hole is affected by frame dragging, namely, the Einstein-Lense-Thirring effect \cite{lense1918einfluss}. This is well known as the Bardeen-Peterson effect\cite{bardeen1975lense}. 
We suggest the coupling of the Blanford-Payne process to the Bardeen-Peterson effect. The concurrence of these two theories implies that the Blanford-Payne process can be distinguished as a theory through the Bardeen-Peterson effect.

\section{Production of AGN Jet outflow.}
\subsection{Blandford-Payne process}
As we remarked in the introduction, the Blanford-Payne process explains the production of AGN jet outflows by employing magnetohydrodynamics (MHD) in 1982. According to the theory, a centrifugally driven outflow of matter from the disk is possible if the poloidal component of the magnetic field makes an angle of less than 60° with the disk surface. At large distances from the disk, the toroidal component of the magnetic field becomes important and collimates the outflow into a pair of antiparallel jets moving perpendicular to the disk. In this way, magnetic stress can extract energy and angular momentum from an accretion disk independently of the presence of viscosity.\cite{blandford1982hydromagnetic}.\cite{https://doi.org/10.48550/arxiv.1508.05192}. Moreover General relativistic magnetohydrodynamics(GRMHD) simulations have shifted the consensus towards the understanding that the Blandford\&Payne mechanism launches sub-relativistic winds that collimate a relativistic jet powered by the BZ-mechanism.\cite{koide_shibata_kudoh_1998},\cite{mizuno_2022}\cite{10.1007/978-94-011-4780-4_66}\cite{ART000959830}

\subsection{Timescale for Accretion jet flow.}
As a simple estimate, the time scale of the whole outflow process, $\Delta t_{BP}$, would be given by;

\begin{equation}
\Delta t_{BP} \sim \frac{l}{c_v}
\end{equation}

Where we denote that $c_v$ is the Alfvén velocity, and $l$ is the length scale over which the poloidal magnetic field extends, which are given as:
\begin{equation}
c_v=(\frac{dp}{d\rho})^{\frac{1}{2}} = \frac{B}{\sqrt{\mu_0\rho}} , 
\end{equation}

\begin{equation}
l=10^3\frac{GM}{c^2}
\end{equation}

Respectively, for the radio jet (or accretion disk) length scale of AGN, where we denote that $\rho$ is the plasma density, $B$ is the magnetic flux, $p$ is the gas pressure of plasma, $\mu_0$ is the vacuum permeability, $G$ is Newton's gravitational constant, $c$ is the speed of light, and $M$ is the mass of the black hole.

\section{Precession of the accretion disk.}
\subsection{The Bardeen-Petterson effect}

The deformation of the accretion disk has been demonstrated  by General relativistic magnetohydrodynamics(GRMHD)\cite{RN42}. the Bardeen-Petterson effect presented the influence on tilted accretion disks around Kerr black holes as astrophysical evidence for frame dragging (Lense-Thirring effect) \cite{bardeen1975lense}.This effect occurs within the Bardeen-Petterson radius, $R_{BP}$, which is given, roughly, by;

\begin{equation}
R_{BP}\gtrsim10^2\frac{GM}{c^2}
\end{equation}

which reaches far outside the horizon.

\subsection{Time scale for accretion disk precession}

Since the total torque involved in the Bardeen-Petterson effect consists of three parts: alignment, precession, and spin-down components, the precession time scale and the alignment time scale would be of the same order;

\begin{equation}
\Delta t_{LT}\sim\frac{2\pi}{\Omega_{GM}}
\end{equation}

Where $\Omega_{GM}$ is the precession angular velocity, which due to the Lense-Thirring effect is given by Wilkins.\cite{wilkins1972bound}

\begin{equation}
\Omega_{GM}= \frac{2GJ}{c^2R^3_{BP}}\lesssim a_*\frac{2G^2M^2}{c^3(100GM/c^2)^3} = 10^{-6}\frac{2c^3a_*}{GM};
\end{equation}

$J$ is the total angular momentum of the black hole, which is given by $J = 2GM^2a_/c$, and $a_*$ is the angular momentum ratio with $0\leq a_* <1$.  \cite{narayan2012observational}.

\section{Luminosity}
Luminous AGN jets are some of the most energetic and powerful phenomena in the universe, and they can produce high-energy radiation and particles that can be detected across the electromagnetic spectrum, from radio waves to gamma rays. AGN jets are classified into two main types based on their luminosity: "quasar-like" jets, which are more luminous, and "radio galaxy" jets, which are less luminous. Quasar-like jets are typically associated with more massive black holes and are more common in the early universe, while radio galaxy jets are typically associated with less massive black holes and are more common in the present-day universe.
AGN jets are thought to be powered by the accretion of matter onto a supermassive black hole at the center of a galaxy. As material falls onto the black hole, it is heated to extremely high temperatures and emits high-energy radiation and particles. Some of this material is then accelerated to near the speed of light and ejected from the vicinity of the black hole in the form of a jet. The particles in the jet are primarily electrons, protons, and other charged particles, and they can be accelerated to energies of up to $10^{20} eV$ or more\cite{blandford1982hydromagnetic}. The jet is also accompanied by a magnetic field, which is thought to play a key role in the acceleration and emission of particles from the jet.

Black hole jets are observed at different frequencies, depending on the specific study and type of jet being observed. The Atacama Large Millimeter Array (ALMA) is an interferometric array of radio telescopes that operate at millimeter and submillimeter wavelengths, corresponding to frequencies that are much higher than those observed by traditional radio telescopes. These frequencies typically range from 30 to 950 GHz. This high-frequency range allows ALMA to make very high-resolution observations of the jets, providing detailed information about their structure and dynamics. For example, ALMA used a 1.3mm (Band 6) wavelength to observe the jet of M87, as shown in (FIG 1) of the study.\cite{Goddi_2021} It is thought that jets from other black holes could be observed using similar methods.

\section{Concurrence of the Blanford-Payne process with the Bardeen-Peterson effect.}

Close to the disk, the plasma flow is driven by gas pressure in a hot, magnetically dominated corona \cite{blandford1982hydromagnetic}. This produces the jet outflow. The disk should be shifted periodically by frame dragging \cite{bardeen1975lense}. Therefore, when the Blanford-Payne process and the Bardeen-Peterson effect work simultaneously, the AGN jet outflow is likely to rotate along the time scale ($T_{LT}$).
 
 We begin with a necessary condition for the process to occur. To make sure that the events can be well resolved in a time sequence, we demand the following condition in this context:

If $\Delta t_{BP} \lesssim \Delta t_{LT}$, the concurrence of the Blandford-Payne process and the Bardeen-Petterson effect can be resolved (Figure 1). This result can be applied to the supermassive black hole (SMBH) the M87, which has been observed directly by the Event Horizon Telescope (EHT). \cite{event2019first}\cite{algaba2021broadband}. SMBH M87 given $M\sim (6.5 \pm  0.7)\times 10^9 M_\odot$\cite{walsh2013m87} and $B=1 \sim 34G$ \cite{akiyama2021first} \cite{akiyama2021first2}. The calculated time scales are $\Delta t_{BP}\sim 1.015$ years and $\Delta t_{LT}\sim 3986.7$ years, which is an impossible value for observation. This is why no one has detected the jet precessions of M87. These results imply that the observable properties can be calculated by selecting smaller black holes. 
In the same manner, Sagittarius $A^*$ (which was directly observed by EHT) has a mass of $M \sim 4\times 10^6 M_\odot$ \cite{akiyama2022first}, The time scales for this black hole are $\Delta t_{BP} \sim 35.6$ hours and $\Delta t_{LT} \sim 15.9$ years. 

We acknowledge that the jet of Sagittarius A* is generally believed to have a relatively low observable power compared to other known black hole jets. But Regarding Sgr A*, recent observations have revealed that it undergoes occasional flares in the submillimeter and infrared bands\cite{murchikova_witzel_2021}, indicating the presence of a compact and variable emission region close to the black hole. While no jet has been directly observed from Sgr A*, it is still a viable candidate for exhibiting jet precession, as the same physical processes that drive jet precession in other systems could also operate in Sgr A*. Furthermore, recent theoretical models have suggested that a Blandford-Znajek powered jet may be present in Sgr A*, which could be relevant to our study.

In our paper, we discuss the timescale for jet precession in the context of accreting black holes, regardless of whether or not they exhibit an observable jet. Our analysis combines well-established theoretical frameworks to provide a quantitative estimate of this timescale, which could help guide future observational campaigns. We acknowledge that there are uncertainties in the existence of Blandford-Payne winds and Bardeen-Petterson alignment, as well as in the applicability of our study to specific objects. However, we believe that our study provides a valuable contribution to the theoretical understanding of jet precession in accreting systems, and we hope that it will stimulate further discussions and investigations in this area.

If the mass of the black hole is smaller, the time scales will be more reliable values to observe. Therefore, we suggest time series observations of black hole jets. If the time series continuity of the concurrence of these two events is  observed, it will have a shape  (FIG 2).

\section{Discussion}
We are discussing the relationship between the Blanford-Payne process and the Bardeen-Peterson effect in the context of AGN jet outflow. We note that when these two processes occur simultaneously, the jet outflow is expected to rotate on a time scale referred to as $T_{LT}$. In order to observe this rotation, We impose a necessary condition that $\Delta t_{BP} \lesssim \Delta t_{LT}$.

We apply this condition to two supermassive black holes, M87 and Sagittarius $A^*$, which have been observed by the Event Horizon Telescope. We find that the calculated time scales for M87 make it impossible to observe the precession of its jet, but the time scales for Sagittarius $A^*$ are more feasible for observation. Based on this, the authors suggest that future observations of black hole jets be made on smaller black holes, where the time scales are more reliable.

Our overall aim is to provide robust support for the Blanford-Payne mechanism and its effectiveness in real-world scenarios. We argue that the inclusion of the Bardeen-Peterson effect is critical in their assessment of the viability of the Blanford-Payne mechanism, as it enables them to accurately evaluate its practical relevance.

In conclusion, We provides a theoretical analysis of the relationship between the Blanford-Payne process and the Bardeen-Peterson effect, and its potential impact on the observation of AGN jet outflows. We suggest that future observations be made on smaller black holes to obtain reliable results, and argue that the inclusion of the Bardeen-Peterson effect is necessary to assess the viability of the Blanford-Payne mechanism.

\section*{Acknowledgement} This research was supported by Basic Science Research Program through the National Research Foundation of Korea(NRF) funded by the Ministry of Education (NRF-2022R1A2C1092602).

\bibliographystyle{plain}
\bibliography{new}

\newpage

\begin{figure}
\centering
\includegraphics[width=0.5 \textwidth]{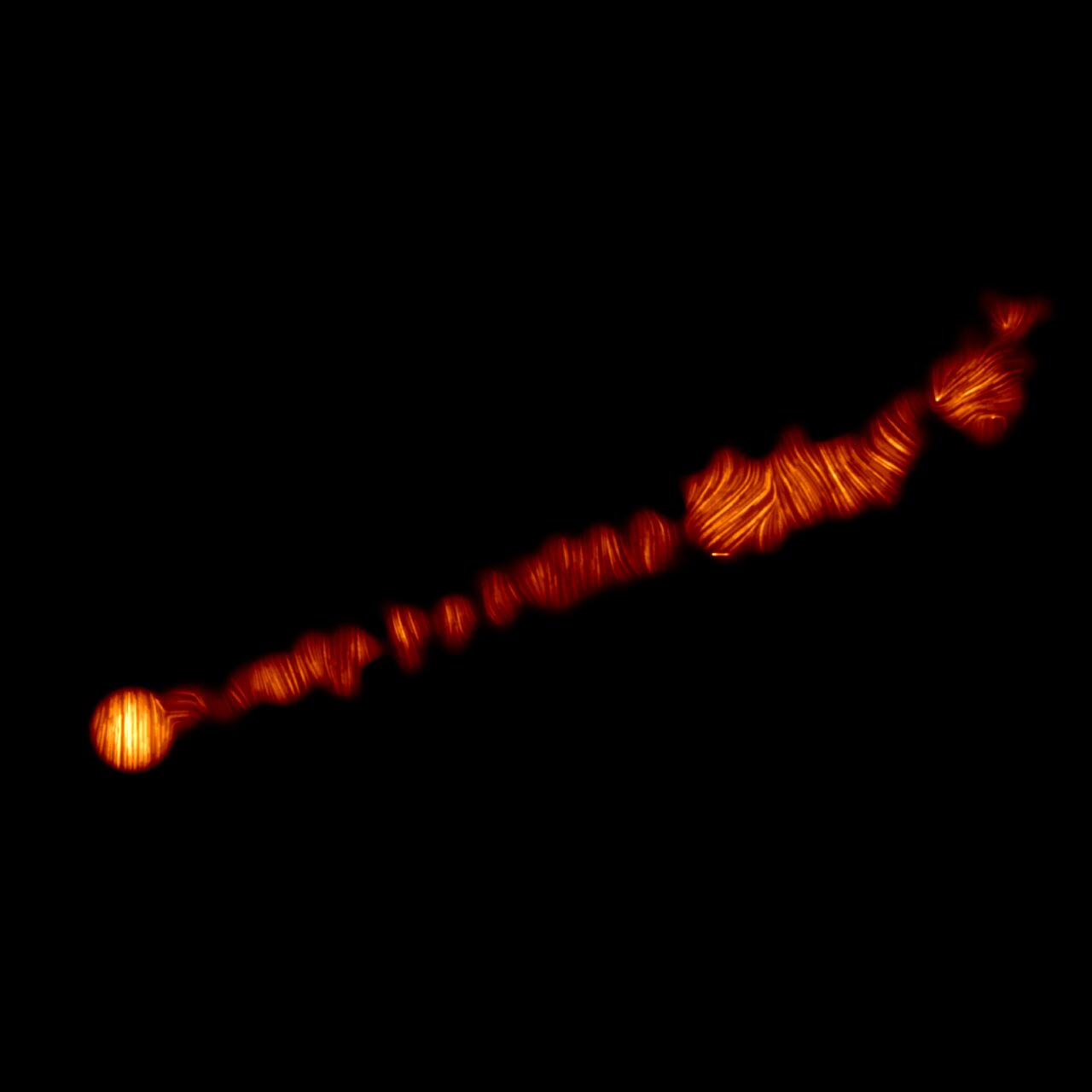}
\caption{polarized light of the M87 jet observed by ALMA. \cite{Goddi_2021}}
\end{figure} 

\begin{figure}
\centering
\includegraphics[width=1 \textwidth]{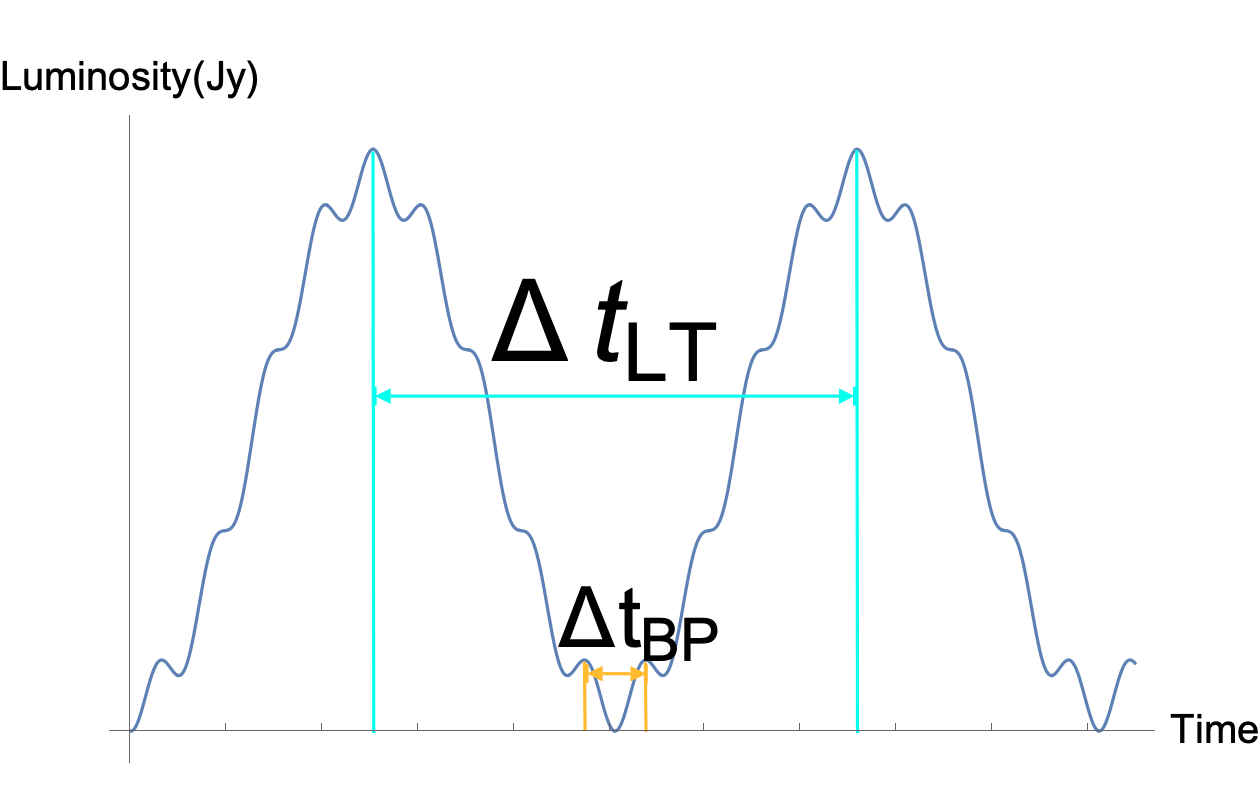}
\caption{Schematic Light Curve of the coupled Blandford Payne process and the Bardeen Petterson effect.}
\end{figure}


\end{document}